\documentclass[aps,prb,floats,amssymb,twocolumn,letterpaper,groupedaddress,
showpacs,floatfix,twoside]{revtex4}
\usepackage{graphicx}
\usepackage{amssymb,amsmath}
\usepackage{epsfig}
\begin{document}

%\preprint{}

\title{
The Effects of Next-Nearest-Neighbor Interactions on the Orientation 
Dependence of Step Stiffness: 
%Bridging the Gap between 
Reconciling  Theory with Experiment for Cu(001)
}


\author{T. J. Stasevich}
\email[]{tjs@glue.umd.edu}
\author{T. L. Einstein}
\thanks{Corresponding author}
\email[]{einstein@umd.edu}
\homepage[]{http://www2.physics.umd.edu/~einstein/}
%\thanks{}
\affiliation{Department of Physics, University of Maryland, College Park, Maryland 20742-4111}
\author{R. K. P. Zia}
\affiliation{Department of Physics, Virginia Polytechnic Institute and State 
University, Blacksburg, Virginia 24601}
\author{M. Giesen}
\author{H. Ibach}
\affiliation{Institut f\"ur Schichten und Grenzfl\"achen, ISG 3,
Forschungszentrum J\"ulich, D 52425 J\"ulich, Germany}
\author{F. Szalma}
\affiliation{Department of Physics, University of Maryland, College Park, MD 20742-4111}

\date{\today}

\begin{abstract}
Within the solid-on-solid (SOS) approximation, we carry out a calculation of the orientational dependence of the step stiffness on a square lattice with nearest and next-nearest neighbor interactions.  At low temperature our result reduces to a simple, transparent expression.  The effect of the strongest trio (three-site, non pairwise) interaction can easily be incorporated by modifying the interpretation of the two pairwise energies.
The work is motivated by a calculation based on nearest neighbors that underestimates the stiffness by a factor of 4 in directions away from close-packed directions, and a subsequent estimate of the stiffness in the two high-symmetry directions alone that suggested that inclusion of next-nearest-neighbor attractions could fully explain the discrepancy.  As in these earlier papers, the discussion focuses on Cu(001).
\end{abstract}

%\keywords{}

\maketitle

\section{Introduction}

At the nano-scale, steps play a crucial role in the dynamics of surfaces.
Understanding step behavior is therefore essential before nano-structures can
be self-assembled and controlled.  
In turn, step stiffness plays a central role in our understanding of
how steps respond to fluctuations and driving forces.  It is one of the three parameters of the step-continuum model,\cite{JW} which has proved a powerful way to describe step behavior
on a coarse-grained level, without recourse to a myriad of
microscopic energies and rates.  As the inertial term, stiffness
determines how a step responds to interactions with other steps,
to atomistic mass-transport processes, and to external driving
forces.  Accordingly, a thorough understanding of stiffness and its
consequences is crucial.

The step stiffness $\tilde{\beta}$ weights deviations from straightness in the step Hamiltonian. Thus, it varies inversely with the step diffusivity, which measures the degree of wandering of a step perpendicular to its mean direction.
This diffusivity can be readily written down in terms of the 
energies $\varepsilon_k$ of kinks
along steps with a mean orientation along close-packed directions ($\langle 110 \rangle$ for an fcc (001) surface): in this case, all kinks are thermally excited.  Conversely, experimental measurements of the low-temperature diffusivity (via the scale factor of the spatial correlation function) can be used to deduce the kink energy.   A more subtle question is how this stiffness depends on the azimuthal misorientation angle, conventionally called $\theta$ and measured from the close-packed direction.  In contrast to $\theta=0$ steps, even for temperatures much below $\varepsilon_k$, there are always a non-vanishing number of kinks, the density of which are fixed by geometry (and so are proportional to $\tan \theta$).  In a bond-counting model, the energetic portion of the step free energy per length (or, equivalently, the line tension, since the surface is maintained at constant [zero] charge\cite{IS}) $\beta(\theta)$ is cancelled by its second derivative with respect to $\theta$, so that the stiffness is due to the entropy contribution alone.  Away from close-packed directions, this entropy can be determined by simple combinatoric factors at low temperature $T$.\cite{rottman,avron,Cahn}  

Interest in this whole issue has been piqued by the recent finding by Dieluweit {\it et al.} \cite{dieluweit02} that the stiffness as predicted in the above fashion, assuming that only nearest-neighbor (NN) interactions $\epsilon_1$ are important, underestimates the values for Cu(001) derived from two independent types of experiments: direct measurement of the diffusivity on vicinal Cu surfaces with various tilts and examination of the shape of (single-layer) islands.  The agreement of the two types of measurements assures that the underestimate is not an anomaly due to step-step interactions.  In that work, the effect of next-nearest-neighbor (NNN) interactions $\epsilon_2$ was crudely estimated by examining a general formula obtained by Akutsu and Akutsu,\cite{AA} showing a correction of order $\exp(-\epsilon_2/k_BT)$, which was glibly deemed to be insignificant.  In subsequent work the Twente group \cite{ZP1} considered steps in just the two principal directions and showed that 
if one included an attractive NNN interaction, one could evaluate the step free energies and obtain a ratio consistent with the experimental results in Ref. \onlinecite{dieluweit02}.  This group later extended their calculations\cite{ZP2} to examine the stiffness.

To make contact with experiment, one typically first gauges the diffusivity along a close-packed direction and from it extracts the ratio of the elementary kink energy $\varepsilon_k$ to $T$. Arguably the least ambiguous way to relate $\varepsilon_k$ to bonds in a lattice gas model is to extract an atom from the edge and place it alongside the step well away from the new unit indentation, thereby creating four kinks.\cite{nelson}  The removal of the step atom costs energy $3\epsilon_1\! +\! 2\epsilon_2$ while its replacement next to the step recoups $\epsilon_1\! +\! 2\epsilon_2$.  Thus, whether or not there are NNN interactions, we identify $\varepsilon_k \! =\! -\scriptstyle{\frac{1}{2}}\textstyle\epsilon_1 \! =\! \scriptstyle{\frac{1}{2}}\textstyle|\epsilon_1|$ (since the formation of Cu islands implies $\epsilon_1 \! <\! 0$); thus, as necessary, $\varepsilon_k \! >\! 0$. 
Note that for clarity we reserve the character $\epsilon$ for lattice-gas energies,\cite{LG} which are deduced by fitting this model to energies which can be measured, such as $\varepsilon_k$.

%Although the kink energy is independent of $\epsilon_2$, 
%in arbitrary directions 
%the NNN interactions does play a significant role in both energy and entropy.

%It turns out that the role of NNN interactions is 
%far more significant than this oversimplified picture initially suggests. 
%%\! + \! 2\epsilon_2 = \epsilon_1 (1 + 2\epsilon_2/\epsilon_1)$.  Thus, an estimate for $\epsilon_1$ from $\varepsilon_k$ differs from the first case by a factor of $(1 \! + \! 2\epsilon_2/\epsilon_1)$.  We shall see that the NNN interaction does significantly more than rescale the effective NN interaction.

The goal of this paper is to compute the step line tension $\beta$ and the stiffness $\tilde{\beta}$ as functions of azimuthal misorientation $\theta$, when NNN (in addition to NN) interactions contribute.  Since it is difficult to generalize the low-temperature expansion of the Ising model,\cite{rottman,avron} we instead study the SOS (solid-on-solid) model, which behaves very similarly at low temperatures and at azimuthal misorientations that are not too large, but can be analyzed exactly even with NNN interactions.  This derivation is described in Section II, with most of the calculational details placed in the Appendix.  In Section III we derive a simple expression for the stiffness in the low-temperature limit, presented in Eq.~(\ref{eq:nnnstiffness}).  We also make contact with parameters relevant to Cu(001), for which this limit is appropriate.  In Section IV we extend the formalism to encompass the presumably-strongest trio (3-atom, non-pairwise) interaction, showing that its effect can be taken into account by shifting the pair energies in the preceding work.  The final section offers discussion and conclusions.

\section{NNN SOS Model on a Square Lattice}

Including NNN interactions in the low-temperature expansion of the square-lattice Ising model lifts the remarkable degeneracy of the model with just NN bonds.  In that simple case, the energy of a path depends solely on the number of NN links, independent of the arrangement of kinks along it; thus, the energy of the ground state is proportional to the number of NN links of the shortest path between two points, and the entropy is related to the number of combinations of horizontal and vertical links that can connect the points.\cite{rottman,Cahn} Including NNN interactions causes the step energy to become a function of both the length of the step and the number of its kinks, eliminating the simple path-counting result.\cite{Cahn}
It can then become energetically favorable for the step to lengthen rather than add another kink.  This causes
the NN energy levels to split in a non-trivial way, making it possible for a longer step to have a lower energy than a shorter step. 
A related complication is that the expansion itself depends on the relative strength of the NNN-interaction:
Instead of an expansion just
in terms of $\exp(-|\epsilon_1|/k_BT)$, the expansion also is in terms of $\exp(\epsilon_2 /2k_BT)$.
Hence, to take the NNN-expansion to the same order of magnitude as the NN-expansion, an unspecified
number of terms is required, depending on the size of the ratio $\epsilon_2/\epsilon_1$.

Since the NNN Ising model cannot be solved exactly and we cannot generalize the low-$T$ expansion, we turn to an SOS model, which was used in earlier examinations of step problems, most notably in the seminal work of Burton, Cabrera, and Frank,\cite{BCF} and later used for steps of arbitrary orientation by Leamy, Gilmer, and Jackson.\cite{Leamy} It was also applied to an interface of arbitrary orientation in a square-lattice Ising model.\cite{Burk}

Although the SOS model can be treated exactly, the result is somewhat unwieldy.  Fortunately, at low temperature---the appropriate regime for the experiments under consideration---the solution reduces to a simple expression.

\subsection{Description of Model}

\begin{figure}[b]
 \includegraphics[angle=0, width=8 cm]{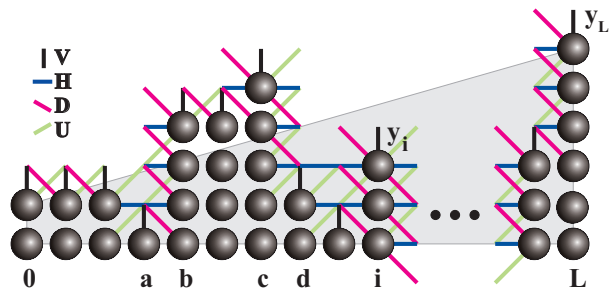}%
 \caption{
A finite-sized step edge whose projected length is $L$.  The step has height
$y_i$ at position $i$ ($0 \leq i \leq L$).  The height difference $y_L - y_0$ is fixed; thus, the step 
edge makes an angle $\theta$ with the horizontal axis, and has an overall slope $m$ (shown as the top of the gray region).  The energy of the step edge is found by counting the number
of broken links required to form it.  Here all NN and NNN broken links are
shown.
}
{\label{fig:nnnbonds}}
 \end{figure}

Consider a step edge of projected length $L$
separating an upper adatom-free region from a lower adatom-filled region (see 
Fig.~\ref{fig:nnnbonds}).
%%%%%%%%%%%%%%%%%%%% FIGURE
%\begin{turnpage}
%\end{turnpage}
%%%%%%%%%%%%%%%%%%%%
The step edge
is completely described by specifying its height $y_i$ at position $i$ ($0 \leq i \leq L$).  The energy of
the step edge depends on the number of broken bonds required to form it.  Let $V$ and $H$ represent the vertical
and horizontal NN bond strengths divided by $k_B T$, and let $U$ and $D$ represent
up-diagonal and down-diagonal NNN bond strengths over $k_B T$.  Then the step-edge energy $E \equiv E(\{\Delta_i\})$
depends only on
$\Delta_i \equiv y_i-y_{i-1}$.

For clarity, we consider two examples.
First, if $\Delta_i=3$ (as is the case between columns a and b in Fig.~\ref{fig:nnnbonds}), 
then between positions $i$ and $i+1$ there are $3$ broken $H$-links,
$2$ broken $U$-links, and $4$ broken $D$-links.  There are also $2$ broken $V$-links, but this number
is independent of $\Delta_i$, since every step-edge configuration of projected length $L$ requires exactly $L$ broken
$V$-links.  Similarly, if $\Delta_i=-3$ (as is the case between columns c and d in 
Fig.~\ref{fig:nnnbonds}), 
then there would be the same number of broken $H$-links, but there
would now be $4$ broken $U$-links and $2$ broken $D$-links (that is, the number of broken $U$ and $D$ links
switch from the previous case).  From these examples we see that, in general, there are $\left| \Delta_i \right|$ broken
$H$-links, $\left| \Delta_i - 1 \right|$ broken $U$-links, and $\left| \Delta_i +1 \right|$ broken $D$-links.
It therefore follows that the step-edge energy is
\begin{eqnarray}
  \label{eq:interfaceEnergy}
\frac{E(\{\Delta_i\})}{k_B T} &=&
\sum_{i=1}^L \Big(V + H\left| \Delta_i \right| + U\left| \Delta_i\! - \! 1\right| \nonumber + D\left| \Delta_i \! + \! 1\right| \Big)\\
&\equiv&
\sum_{i=1}^L K(\Delta_i).
\end{eqnarray}

Because we seek the orientation dependence of $\beta$ and $\tilde{\beta}$, we constrain the step to have an overall
offset $Y \equiv y_L - y_0 \equiv L \tan \theta = \sum_{i=1}^L \Delta_i$. (This constraint is represented in 
Fig.~\ref{fig:nnnbonds} by the shaded gray area.  Equivalently, we specify that the overall slope of the step is $m \equiv \tan \theta$.)  The constrained partition function is
 therefore
\begin{eqnarray}
  \label{eq:partfunction}
  Z\left( Y\right) \equiv \sum_{\left\{ \Delta \right\} }\delta \left[
Y-\sum_{i=1}^L\Delta _i\right] e^{-E(\{\Delta_i\})/k_B T},
\end{eqnarray}
\noindent where $\{\Delta\}$ is the set of all $\Delta_i$ each of which ranges over all integers.  From $Z(Y)$ we can find the orientation dependence of the
free energy $F(Y) = -k_B T \ln Z(Y)$, the {\it projected} free energy per length 
$f(m) = F(Y)/L$, and the line tension (or free energy per length)
$\beta(\theta) = f(m)\cos \theta$ (since the step length is $L/ \cos \theta$);
thence, we can find the stiffness $\tilde{\beta}(\theta) =\beta (\theta )+
\partial^2\beta (\theta )/\partial \theta ^2$.

For future reference, note that the process of extracting an atom from the step-edge and replacing it alongside the edge, discussed in the penultimate paragraph of the introduction, creates two pairs of $\Delta \! =\! +1$ and $\Delta \! =\! -1$, costing $4H$ according to Eq.~(\ref{eq:interfaceEnergy}) and removing a net of 2 NN bonds, so that $H \! =\! -\epsilon_1/2k_BT\! =\! \varepsilon_k/k_BT$.  Similarly, we compare the energies of two NN atoms, abutting [the lower side of] a step edge ($\{\Delta_{i}\} \! =\! 0$) at $i_0$ and either parallel or perpendicular to the edge.  In the first case, $\Delta_{i_0} \! =\! +1$ and $\Delta_{i_0+2} \! =\! -1$, with an added energy of $2H + 2(U\! +\!D)$ according to Eq.~(\ref{eq:interfaceEnergy}). In the perpendicular case $\Delta_{i_0} \! =\! +2$ and $\Delta_{i_0+1} \! =\! -2$, implying an added energy of $4H + 4(U\! +\!D)$.  Counting bonds we see that the parallel configuration has one more $\epsilon_1$ bond and two more $\epsilon_2$ bonds than the perpendicular configuration. Invoking $H \! =\! -\epsilon_1/2k_BT$, we see that $U\! +\!D \! =\! -\epsilon_2/k_BT$; if $U\! =\!D$, then $D \! =\! -\epsilon_2/2k_BT$.  The factor-of-2 difference between broken links in Eq.~(\ref{eq:interfaceEnergy}) and broken bonds was noted (for H links) already in the classic exposition by Leamy et al.\cite{Leamy} An alternate argument, presented over a decade ago,\cite{ZELD} for this factor of 2 is that the ragged edge is created by severing bonds along the selected path through an infinite square.  This leads to the formation of two complementary irregular boundary layers (with opposite values of $\{\Delta_{i}\}$, so that the associated energy of each is half that of the broken bonds (at least when $U\! =\!D$).  

\subsection{Evaluation of the Free Energy}

As detailed in the first part of the Appendix, the sum in the Fourier
transform of $Z(Y)$, which we denote by $W(\mu )$, factorizes. Thus, it can
be written as 
\[
W(\mu )=\exp \left[ -Lg(i\mu )/k_BT\right] , 
\]
where $g(i\mu )$ is the reduced Gibbs free energy per column. To evaluate
the inverse transform, we exploit the saddle point method and obtain (see 
Appendix for details)

\begin{equation}
Z(Y)\approx \exp \left[ -L\left( \rho _0\tan \theta +\frac{g(\rho _0)}{k_BT}%
\right) \right] ,  \label{eq:ZtoWapproximation}
\end{equation}
where the saddle point ($\mu_0 =$ $-i\rho _0$) is defined implicitly by the
stationarity condition 
\begin{equation}
-\frac{g^{\prime }(\rho _0)}{k_BT}=m\equiv \tan \theta .
\label{eq:saddlerho}
\end{equation}
\noindent Here, prime (as in $g^{\prime }$) denotes a derivative with respect
to $\rho $. This result can be regarded as applying a ``torque'' to the step
to produce a rotation $\theta =\tan ^{-1}m$ from the minimum-energy,
close-packed orientation.\cite{Leamy}

Taking the logarithm of Eq.~(\ref{eq:ZtoWapproximation}), we find the
projected free energy per column $f(m)$ as a Legendre transform of the
reduced Gibbs free energy per column $g(\rho _0)$: 
\begin{equation}
\frac{f(m)}{k_BT}\approx \rho _0m+\frac{g(\rho _0)}{k_BT}.  \label{eq:ftog}
\end{equation}
Note that this expression is valid only for $L\gg 1$; for finite-sized
systems, corrections are required. As standard for Legendre transforms,\cite{KSK} we
have 
\begin{equation}
\frac{\ddot{f}(m)}{k_BT}=-\frac{k_BT}{g^{\prime \prime }(\rho _0)},
\end{equation}
where $\ddot{f}\equiv \partial^2 f/\partial m^2$. Using $\beta (\theta)a=f(m )\cos
\theta $ and $m=\tan \theta $, with $a$ the lattice constant of the square (i.e., the column spacing, which is $(1/\sqrt{2})$ the conventional fcc lattice constant), we can rewrite the stiffness as 
\begin{equation}
\tilde{\beta}(\theta)a=\ddot{f}(m)/\cos ^3\theta ,  \label{fstiff}
\end{equation}
or, similar to results by Bartelt \textit{et al}.,\cite{BEW} 
\begin{equation}
\frac{k_BT}{\tilde{\beta}(\theta )a}=-\frac{g^{\prime \prime }(\rho _0)}{k_BT}%
\cos ^3\theta .  \label{eq:gtostiffrho}
\end{equation}
Thus, we only need $g^{\prime \prime }(\rho )$ to find the stiffness as a
function of $m$ or $\theta $.

Of course, $\rho _0$ in $g^{\prime \prime }$ must be eliminated in favor of $%
m$ via Eq.~(\ref{eq:saddlerho}). The details for the general case are
somewhat involved. Here, we simplify to the physically relevant case of $U=D$
and, defining $S\equiv H+U+D=H+2D$, just quote the results: 
\begin{equation}
\frac{g^{\prime \prime }\left( \rho _0\right) }{k_BT}=-m\left[ \ \frac{%
2\sinh \rho _0}{C(S,\rho_0)}+\coth \rho _0\right] \ +m^2
\label{eq:ddgfinal}
\end{equation}
where $C(S,\rho_0) \equiv \cosh S -\cosh \rho_0$ and $\rho _0(m)$ is found by inverting
\begin{equation}
\hspace{-.3in}m=\frac{\sinh \rho _0\sinh S}{C(S,\rho_0)\left[ \sinh S-C(S,\rho_0) \left( 1-e^{-2D}\right) \right] }.
\label{eq:rhotom}
\end{equation}
Some details can be found in the Appendix. Since Eq.~(\ref{eq:rhotom}) is a
quartic equation for $\cosh \rho _0$ or $e^{\rho _0}$, the explicit
expression for $\rho _0(m)$ is rather opaque. However, at low-temperatures,
a simpler formula emerges, as shown in the next section.

%\vspace{20 pt}

\section{Low-T Solution: Simple Expression}

At low temperatures, we find that the appropriate root for $\rho _0$
diverges. Then we can write $\cosh \rho _0\approx \sinh \rho _0\approx
e^{\rho _0}/2$. Of course, $H\propto 1/T$ so that $\cosh S\approx e^S/2$.
With these approximations, Eq.~(\ref{eq:rhotom}) becomes quadratic in $%
e^{\rho _0}$: 
\begin{eqnarray}
  \label{eq:lowT}
  m = \frac{ e^{\rho_0 + S}  }{(e^S-e^{\rho_0})[e^S-(e^S-e^{\rho_0})(1-e^{-2 D})]}
\end{eqnarray}
Likewise, the expression for $g^{\prime \prime}(\rho_0)$, Eq.~(\ref{eq:ddgfinal}), becomes
\begin{eqnarray}
  \label{eq:lowTddg}
\frac{g^{\prime \prime }\left( \rho_0 \right)}{k_B T}
&=&-m \left[\ \frac{2 e^{\rho_0}}{(e^S - e^{\rho_0})} + 1 \right]\ + m^2.
\end{eqnarray}
Solving for $e^{\rho_0}$ in Eq.~(\ref{eq:lowT}) and inserting the solution into Eq.~(\ref{eq:lowTddg}) gives
\begin{eqnarray}
   \label{eq:ddglowt}
\frac{g^{\prime \prime }\left( \rho_0\right)}{k_B T}
&=&-m  \sqrt{(1 - m)^2 + 4 m e^{-2D}}.
\end{eqnarray}
so that, from Eq.~(\ref{eq:gtostiffrho}), and recalling $D \! =\! -\epsilon_2/2k_BT$, we arrive at our {\it main result}, a simple, algebraic expression for
$\tilde{\beta}$ as a function of $m$:
\begin{eqnarray}
  \label{eq:nnnstiffness}
  \frac{k_B T}{\tilde{\beta}a} = \frac{m \sqrt{(1-m)^2 + 4me^{\epsilon_2/k_BT}}}
  {( 1 + m^2 )^{3/2}  }.
\end{eqnarray}
We examine Eq.~(\ref{eq:nnnstiffness}) in several different limiting cases.  When $\epsilon_2=0$, this reduces to
\begin{eqnarray}
 \label{eq:stiffd=0}
  \frac{k_B T}{\tilde{\beta}a} = \frac{m + m^2  }{\left( 1 + m^2 \right)^{3/2}  }.
\end{eqnarray}
as found in a previous study involving only NN interactions.\cite{dieluweit02}  Interestingly, 
at $\theta = 45^\circ$, Eq.~(\ref{eq:nnnstiffness}) shows a simple dependence on $\epsilon_2$, namely,
\begin{eqnarray}
  \label{eq:m=1}
     \frac{k_B T}{\tilde{\beta}a}=\frac{e^{\epsilon_2/2k_BT}}{\sqrt{2}}.
\end{eqnarray}
Of course, this reduces to the venerable Ising result of $1/\sqrt{2}$ in the absence of 
NNN interactions ($\epsilon_2 \! =\! 0$).\cite{rottman,Ising45,Z00}

By considering just the lowest and second lowest energy configurations,\cite{ZP1,ZP2} Zandvliet et al.\ obtained the result\cite{ZP2} (expressed with our sign convention for $\epsilon_2$) for the maximally misoriented case $m \! =\!1$

\begin{eqnarray}
  \label{eq:ZVP}
     \frac{k_B T}{\tilde{\beta}a}=\frac{\sqrt{2}}{1 + e^{-\epsilon_2/2k_BT}},
\end{eqnarray}
which has, for the attractive $\epsilon_2$ of primary concern here, some qualitative similarities to Eq.~(\ref{eq:m=1}) (including the value $1/\sqrt{2}$ for $\epsilon_2 \! =\! 0$) but is too small by a factor of 2 for $\epsilon_2/2k_BT \ll 0$; even the coefficient of the first-order term in an expansion in $\epsilon_2/2k_BT$ is half the correct value. For the opposite limit of repulsive $\epsilon_2$, Eq.~(\ref{eq:ZVP}) levels off (at $\sqrt{2}$), in qualitative disagreement with the actual exponential increase seen in Eq.~(\ref{eq:m=1}).
%%%%%%%%%%%%%%%%%%%% FIGURE
%\begin{turnpage}
 \begin{figure}[b]
 \includegraphics[angle=0, width=9 cm]{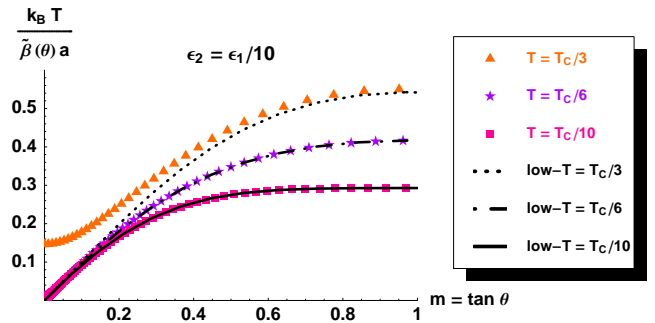}%
 \caption{
The range of validity of Eq.~(\ref{eq:nnnstiffness}) is examined by
comparing it to exact numerical solutions of the SOS model at
several temperatures.  In the legend $T_c$ refers to the NN lattice-gas (Ising) model; for $|\epsilon_1|= 256$meV, $T_c = 1685$K.}
{\label{fig:nnnCompare}}
 \end{figure}
%\end{turnpage}
%%%%%%%%%%%%%%%%%%%%
%%%%%%%%%%%%%%%%%%%% FIGURE
%\begin{turnpage}
 \begin{figure}[b]
 \includegraphics[angle=0, width=9 cm]{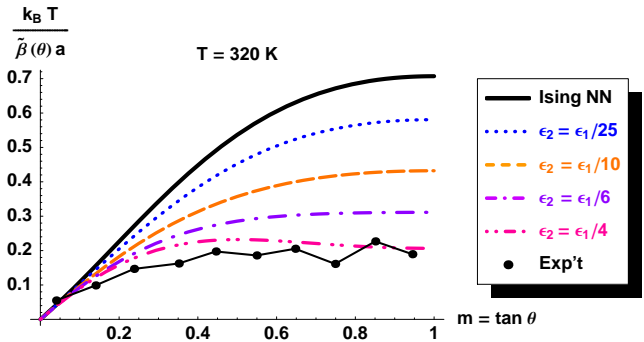}%
 \caption{
Eq.~(\ref{eq:nnnstiffness}) is plotted for a variety of different values of $D = -\epsilon_2/2k_B T$, where
$\epsilon_1$ and $\epsilon_2$ are NN- and NNN-interaction energies, respectively, in a lattice-gas picture.  The solid curve denoted ``Ising NN" corresponds to $\epsilon_2 \! =\! 0$.  The dots labeled ``Exp't" are taken from Fig.~2 of Ref.~\onlinecite{dieluweit02} and were derived from the equilibrium shape of islands on Cu(001) at 302K, with the line segments serving as guides for the eye.  To minimize clutter, we omit similar data derived from correlation functions of vicinal surfaces at various temperatures.  Note that for $\epsilon_2 \! =\! \epsilon_1/4$ a maximum has developed near $\tan \theta \! =\! 1/2$ that is not evident in the experimental data. 
}
{\label{fig:nnnplot}}
 \end{figure}
%\end{turnpage}
%%%%%%%%%%%%%%%%%%%%

Fig.~\ref{fig:nnnCompare} compares Eq.~(\ref{eq:nnnstiffness}) to corresponding exact solutions
[found by numerically solving
Eqs.~(\ref{eq:gtostiffrho}), (\ref{eq:ddgfinal}), and (\ref{eq:rhotom})]
at several temperatures when $\epsilon_2 = \epsilon_1/10$.
We see that Eq.~(\ref{eq:nnnstiffness}) overlaps the exact solution at temperatures as high as $T_c/6$.
As the temperature increases, the stiffness becomes
more isotropic, and Eq.~(\ref{eq:nnnstiffness}) begins to overestimate the stiffness 
near $\theta = 0^\circ$.

Finally, in Fig.~\ref{fig:nnnplot} 
(using the experimental value\cite{GI} $\varepsilon_k = 128$ meV $\Rightarrow \epsilon_1 = -256$ meV), we compare Eq.~(\ref{eq:nnnstiffness}) to the NN-Ising model at $T=320$K, as well as to the experimental results of Ref.~\onlinecite{dieluweit02}.
For strongly attractive (negative) $\epsilon_2$,
$k_B T / \tilde{\beta}a$ decreases significantly.  In fact, when $\epsilon_2/\epsilon_1$ is 1/6,
so that $-\epsilon_2/2k_BT = (\epsilon_2/\epsilon_1)(\varepsilon_k/k_BT) \approx (1/6)4.64$,
the model-predicted value of $k_B T/ \tilde{\beta}a$ has decreased to less than half its $\epsilon_2 \!=\! 0$ value (viz. by a factor of 0.46, vs. 0.63 if Eq.~(\ref{eq:ZVP}) is used), so about 3/2 the experimental ratio.
If $\epsilon_2/\epsilon_1$ increases even further, $k_B T/ \tilde{\beta}a$
further decreases and develops positive curvature, causing an
endpoint local minimum to appear at $\theta = 45^\circ$.  We can determine when this occurs by expanding Eq.~(\ref{eq:nnnstiffness}) about $m=1$:
\begin{eqnarray}
  \label{eq:mexpand}
  \frac{k_B T}{\tilde{\beta}a} =
%\frac{e^{\frac{\epsilon_2}{2k_BT}}}{\sqrt{2}} + \left( %\frac{e^{-\frac{\epsilon_2}{2k_BT}}}{8 \sqrt{2}} \! -\!\frac{3 %e^{\frac{\epsilon_2}{2k_BT}}}{4 \sqrt{2}}  \right) (m\! -\!1)^2 \! +\! \ldots
\frac{e^{-D}}{\sqrt{2}} + \left( \frac{e^{D}}{8 \sqrt{2}} \! -\!\frac{3 e^{-D}}{4 \sqrt{2}}  \right) (m\! -\!1)^2 \! +\! \ldots
\end{eqnarray}
Setting the coefficient of $(m-1)^2$ to zero gives $-2D = \epsilon_2/k_BT = -\ln(6) \approx -1.8$, which corresponds to a value of
$k_B T/\tilde{\beta}a=\sqrt{3}/6  \approx 0.29$, about 2/5 the value at $\epsilon_2 \! =\! 0$.  For $T=320$K and $\varepsilon_k = 128$ meV, this corresponds to $\epsilon_2/\epsilon_1 \approx 0.2$. However, for the NNN interaction alone to account for the factor-of-4 discrepancy between model/theory and experiment reported by Dieluweit et al.\cite{dieluweit02}, Fig.~\ref{fig:nnnplot} shows that $\epsilon_2/\epsilon_1 \approx 0.3$ would be required.

\section{Effect of Trio Interactions}

In addition to the NNN interaction, trio (3-atom, non-pairwise) interactions may well influence the stiffness.  The strongest such interaction is most likely that associated with 3 atoms forming a right isosceles triangle, whose sides are at NN distance and hypotenuse at NNN separation.  In a lattice gas model, there is a new term with $\epsilon_{RT}$ times the occupation numbers of the 3 sites.\cite{LG}  Note that this trio interaction energy $\epsilon_{RT}$ is in addition to the contribution $2\epsilon_1 \! +\! \epsilon_2$ of the constituent pair interactions.  If we count broken trios and weight each by $R$, we find an additional contribution to Eq.~(\ref{eq:interfaceEnergy}) of $R$ times 
\begin{eqnarray}
  \label{eq:trio}
4|\Delta_i| +2\delta_{\Delta_i,0} +2 = 2|\Delta_i| + |\Delta_i \! +\! 1| + |\Delta_i \! -\! 1| +2,
\end{eqnarray}
where we have converted the Kronecker delta at $i\! =\! 0$ to make better contact with Eq.~(\ref{eq:interfaceEnergy}).  Thus, without further calculation we can include the effect of this trio by replacing $H$ by $H\! +\! 2R$, $U$ by $U\! +\! R$, $D$ by $D\! +\! R$, and (trivially) $V$ by $V\! +\! 2$.  

By arguments used at the end of Section IIA, we recognize $R\! =\! -\scriptstyle{\frac{1}{2}}\textstyle{\epsilon_{RT}}$.  Consequently, the effective NN lattice-gas energy is $\epsilon_1\! +\! 2\epsilon_{RT}$ and, more significantly the effective NNN interaction energy is $\epsilon_2\! +\! \epsilon_{RT}$.  Thus, $\epsilon_{RT}$ must be attractive (negative) if it is to help account for the discrepancy in Fig.~2 of Ref.~\onlinecite{dieluweit02} between the model and experiment.  Furthermore, by revisiting the configurations discussed in the penultimate paragraph of the Introduction, we find that the kink energy $\varepsilon_k$ becomes $-\scriptstyle{\frac{1}{2}}\textstyle{\epsilon_1} -\! \epsilon_{RT}$.  Thus, for a repulsive $\epsilon_{RT}$, $|\epsilon_1|$ will be larger than predicted by an analysis of, e.g., step-edge diffusivity that neglects $\epsilon_{RT}$.
Lastly, the close-packed edge energy, i.e. the $T\! =\! 0$ line tension $\beta(0) \! =\! -\scriptstyle{\frac{1}{2}}\textstyle{\epsilon_1}  -\! \epsilon_2$, becomes 
$-\scriptstyle{\frac{1}{2}}\textstyle{\epsilon_1} - \! \epsilon_2 -\! 2\epsilon_{RT}$

\section{Discussion and Conclusions}

We now turn to experimental information about the interactions, followed by comments on the limited available calculations of them, often recapitulating the discussion in Ref.~\onlinecite{ZP1}.  All the experiments are predicated on the belief that at 320K there is sufficient mobility to allow equilibrium to be achieved.  If the NNN interactions are to explain at least partially the high stiffness of experiment compared to Ising theory, the NNN interaction must be attractive and a substantial fraction of $\epsilon_1$. Since compact islands do form on the Cu(001) surface, it is obvious that $\epsilon_1$ is attractive.  If $\epsilon_2$ is also attractive, as required for reduction of the overestimate of $k_BT/\tilde{\beta}$, then the low-temperature equilibrium shape has clipped corners (octagonal-like, with sides of alternating lengths), as noted in Ref.~\onlinecite{ZP1}; no evidence of such behavior has been seen.  The lack of evidence of a decreasing stiffness near $\theta \approx 45^\circ$ suggests that $\epsilon_2/\epsilon_1$ is at most 1/5.

There is implicit experimental information for $\epsilon_2$: from island shapes\cite{Ising45} and fluctuations\cite{SGVI} $\beta(0) = 220 \pm \! 11\!$ meV.  Since related measurements showed $\scriptstyle{\frac{1}{2}}\textstyle{\epsilon_1} \!=\! -128$meV, we deduce $\epsilon_2 \!=\! -92 \!$ meV if $\epsilon_{RT}$ is insignificant. These values imply that $\epsilon_2/\epsilon_1$ is somewhat larger than 1/3, which seems unlikely in light of the unobserved predictions about the shape of islands in that case (cf. the end of Section III). 

To corroborate this picture, one should estimate the values of $\epsilon_1$ and $\epsilon_2$, as well as $\epsilon_{RT}$, from first-principles total-energy calculations.  In contrast to Cu(111),\cite{Bogi,Feibel} however, no such information even for $\epsilon_1$ has been published for Cu(001); there are, however, several semiempirical calculations which found $\varepsilon_k \approx 0.14$eV.\cite{semi}   In such calculations based on the embedded atom method (EAM), which work best for late transition and noble fcc metals, the indirect (``through-substrate") interactions are expected to be strong only when the adatoms share common substrate nearest neighbors; then the interaction should be repulsive and proportional to the number of shared substrate atoms.\cite{TLE-Unertl}  (Longer range pair interactions and multisite non-pairwise interactions are generally very-to-negligibly small in such calculations; they probably underestimate the actual values of these interactions since there is no Fermi surface in this picture, and it is the Fermi wavevector that dominates long-range interactions.) If the NN and NNN interactions on Cu(001) were purely indirect, we would then predict $\epsilon_2 \!=\! \scriptstyle{\frac{1}{2}}\textstyle{\epsilon_1} \! > \! 0$.  However, whenever direct interactions (due to covalent effects between the nearby adatoms) are important, they overwhelm the indirect interaction.  At NN separation, which is the
 bulk NN spacing, direct interactions must be significant, explaining why $\epsilon_1$ can be attractive.  It is not obvious from such general arguments whether there are significant direct interactions between Cu adatoms at NNN separations.  (For Pt atoms on Pt(100), the only homoepitaxial case in which $\epsilon_2$ was computed semiempirically, EAM calculations\cite{WDF} gave $\epsilon_2/|\epsilon_1|$ \!=\! 0.2, less than half the  ratio predicted by counting substrate neighbors, but with the predicted repulsive $\epsilon_2$.) It is also not obvious {\it a priori} whether multi-atom interactions also contribute significantly.  (For homoepitaxy, the only semiempirical result is that they are insignificant for Ag on Ag(001);\cite{VA} however, it is likely that semiempirical calculations will underestimate multi-atom interactions.) 

To address these questions, we are currently carrying out calculations\cite{STK} using the VASP package.\cite{VASP}  Preliminary results for Cu(001) suggest that $\epsilon_2$ is indeed attractive, and that $\epsilon_2/\epsilon_1$ is about 1/8; however, there are indications of a {\it repulsive} right-triangle trio interaction $\epsilon_{RT}$ with sizable magnitude (perhaps comparable to $|\epsilon_2|$, consistent with {\it a priori} expectations\cite{TLE-Unertl,TLE-Maine}), which would diminish rather than enhance the effect of $\epsilon_2$.

In summary, NNN interactions may well account for a significant fraction, perhaps even a majority, of the discrepancy between NN Ising model calculations and experimental measurements of the orientation dependence of the reduced stiffness;\cite{dieluweit02} the effect is even somewhat greater than estimated by the Twente group\cite{ZP1,ZP2}.  However, inclusion of $\epsilon_2$ is not the whole answer, nor, seemingly, is consideration of $\epsilon_{RT}$.  One possible missing ingredient is other multi-site interactions, most notably the linear trio $\epsilon_{LT}$ consisting of 3 colinear atoms (a pair of NN legs and an apex angle of $180^{\circ}$).  In a model calculation their energy was comparable to $\epsilon_{RT}$,\cite{TLE-Unertl,TLE-Maine} albeit with half as many occurrences per atom in the monolayer phase.  The corrections due to $\epsilon_{LT}$ would be more complicated than simple shifts in the effective values of $\epsilon_1$ and $\epsilon_2$.
Since direct interactions are probably important, there is no way to escape doing a first-principles computation;
we continue to use the VASP package to extend our
preliminary calculations.\cite{STK}  A more daunting (at least for lattice-gas afficionadoes) possibility is that long-range intrastep elastic effects may be important.  Ciobanu and Shenoy have made noteworthy progress in understanding how this interaction contributes to 
the orientation dependence of noble-metal steps.\cite{shenoy}
 
\appendix*
\section{Calculational Details}
\subsection{Partition Function}

To carry out the sum in Eq. (\ref{eq:partfunction}), we consider the Fourier
transform of $Z(Y)$: 
\begin{eqnarray}
W\left( \mu \right) &\equiv &\int_{-\infty }^\infty dY~e^{i\mu Y}Z\left(
Y\right)  \nonumber  \label{eq:fourier} \\
&=&\sum_{\left\{ \Delta \right\} }\exp \sum_{j=1}^L\left( i\mu \Delta
_j-K\left( \Delta _j\right) \right)  \nonumber \\
&=&\left[ \sum_{\Delta =-\infty }^\infty \exp \left( i\mu \Delta -K\left(
\Delta \right) \right) \right] ^L,
\end{eqnarray}
where $K\left( \Delta \right) \equiv \left( V+H\left| \Delta \right|
+U\left| \Delta -1\right| +D\left| \Delta +1\right| \right) $ is the energy
in Eq. (\ref{eq:interfaceEnergy}), associated with adjacent columns with height difference 
$\Delta $. Carrying out the summation in Eq.~(\ref{eq:fourier}) gives 
\begin{equation}
\frac{g\left( i\mu \right) }{k_BT}\equiv -\frac 1L\ln W(i \mu )=V+U+D-\ln
B(i\mu ),  \label{eq:gdefine}
\end{equation}
where 
\begin{equation}
B(i\mu )\equiv 1+\frac{e^{2D}}{e^{H+U+D+i\mu }-1}+\frac{e^{2U}}{%
e^{H+U+D-i\mu }-1}.  \label{eq:defineB}
\end{equation}
Thus, the original partition function $Z(Y)$ is: 
\begin{eqnarray}
Z\left( Y\right) &=&\frac 1{2\pi }\int_{-\infty }^\infty d\mu ~e^{-i\mu
Y}W\left( \mu \right)  \label{eq:ZtoW} \\
&=&\frac 1{2\pi }\int_{-\infty }^\infty d\mu ~\exp \left[ L\left( -i\mu \tan
\theta -\frac{g(i\mu )}{k_BT}\right) \right]  \nonumber
\end{eqnarray}
For $L\gg 1$, we can evaluate this inverse transform by steepest decent
approximation. The saddle point occurs on the imaginary axis ($\mu
=-i\rho $), at the value $\rho _0$ given by the stationary-phase condition: 
\begin{equation}
-\frac{g^{\prime }\left( \rho _0\right) }{k_BT}=m\equiv \tan \theta . 
\end{equation}
Calculating the derivative from Eqs.~(\ref{eq:gdefine}) and (\ref{eq:defineB}), we find 
\begin{equation}
m=B^{\prime }(\rho _0)/B(\rho _0),  \label{eq:mtorho}
\end{equation}
where prime stands for $\partial _\rho $. The leading contribution to this
integral (\ref{eq:ZtoW}) is just the integrand evaluated at this point: 
\begin{equation}
Z(Y)\approx \exp \left[ -L\left( m\rho _0+\frac{g\left( \rho _0\right) }{k_BT%
}\right) \right] .  \label{eq:ZtoWapproximationmu}
\end{equation}

\subsection{Analysis of $g^{\prime \prime }(\rho )$ and specialization to $%
U=D$}

From Eqs.~(\ref{eq:gdefine}), we find 
\begin{equation}
\frac{g^{\prime }\left( \rho \right) }{k_BT}=-B^{\prime }(\rho )/B(\rho )
\label{eq:dgtoB}
\end{equation}
and 
\begin{equation}
\frac{g^{\prime \prime }\left( \rho \right) }{k_BT}=-B^{\prime \prime }(\rho
)/B(\rho )+\left[ B^{\prime }(\rho )/B(\rho )\right] ^2.  \label{eq:ddgtoB}
\end{equation}
This can be simplified, by Eq.~(\ref{eq:mtorho}), to 
\begin{equation}
\frac{g^{\prime \prime }\left( \rho _0\right) }{k_BT}=-mB^{\prime \prime
}(\rho _0)/B^{\prime }(\rho _0)+m^2,  \label{eq:ddgrho}
\end{equation}
the quantity needed for computing the stiffness as a function of $m$. While
straightforward, computing the derivatives with the general form for $B$
(Eq.~(\ref{eq:defineB}) with $\rho =i\mu $) is quite tedious.  A
slight simplification emerges if we specialize to the physically relevant
case $U=D$. Then, with $S\equiv H+2D$, we have 
\begin{eqnarray}
B(\rho ) &=&1+\frac{e^{2D}}{e^{S+\rho }-1}+\frac{e^{2D}}{e^{S-\rho }-1} 
\nonumber \\
&=&1-e^{2D}+\frac{e^{2D}\sinh S}{\cosh S-\cosh \rho } \nonumber \\
& \equiv &1-e^{2 D}+\frac{e^{2D} \sinh S}{C(S,\rho)},
\end{eqnarray}
so that 
\begin{equation}
B^{\prime }(\rho )=e^{2D}\sinh S\frac{\sinh \rho }{C^2(S,\rho)},
\end{equation}
and 
\begin{equation}
B^{\prime \prime }(\rho )=e^{2D}\sinh S\left[ \frac{\cosh \rho }{%
C^2(S,\rho)}+\frac{2\sinh ^2\rho }{C^3(S,\rho)}%
\right] .
\end{equation}
Inserting these expressions into Eq.~(\ref{eq:mtorho}), we have
\begin{equation}
m=\frac{\sinh \rho _0\sinh S}{C(S,\rho_0)\left[ \sinh S-C(S,\rho_0)\left( 1-e^{-2D}\right) \right] }.
\end{equation}
Similarly, with Eq. (\ref{eq:ddgrho}), we find 
\begin{equation}
\frac{g^{\prime \prime }\left( \rho _0\right) }{k_BT}=-m\left[ \ \frac{%
2\sinh \rho _0}{C(S,\rho_0)}+\coth \rho _0\right] \ +m^2.
\end{equation}

\section*{Acknowledgment}

Work at the University of Maryland was supported by the NSF-MRSEC, Grant DMR 00-80008.  One of us (RKPZ) acknowledges support by NSF Grants DMR 00-88451 and 04-14122.  TLE acknowledges partial support of collaboration with ISG-3 at FZ-J\"ulich via a Humboldt U.S. Senior Scientist Award.
We have benefited from ongoing interactions with E.~D.\ Williams and her group.

%%%%%%%%%%%%%%%%%%%%%%% Bibliography %%%%%%%%%%%%%%%%%%%%%%%%%%%%%%%


\begin{thebibliography}{10}
\bibitem{LG} Explicitly, the contribution to the lattice-gas Hamiltonian of all NN bonds is  $\epsilon_1 \Sigma_{\langle i,j\rangle} n_in_j$, where the site-occupation variable $n_i \! = \! 0,1$, and the summation is over all NN pairs of sites.  It is well known that $\epsilon_1 \rightarrow -4J_1$ in the corresponding Ising model, so that $T_c$ is determined by $\sinh(|\epsilon_1|/2k_BT) \! = \! 1$. Unfortunately, the variety of notations in papers on this subject can lead to confusion.  In Refs.~\onlinecite{ZP1,ZP2}, $\varepsilon_{1,2}$ have the opposite sign of our $\epsilon_{1,2}$.  In Ref.~\onlinecite{Ising45} and somewhat implicitly in Ref.~\onlinecite{dieluweit02}, the so-called the Ising parameter, $\varepsilon$, is $\varepsilon_k \! = \! 2J \! = \! -\scriptstyle{\frac{1}{2}}\textstyle\epsilon_1$.
\end{thebibliography}
\end{document}